\documentstyle[11pt]{article}
\begin{document}
\begin {center}

\bf QSO 0957+561 AND OTHER LARGE-SEPARATED DOUBLE QUASARS: SOME NEW RESALTS
AND A~FUTURE OBSERVATIONAL PROJECT\\

\smallskip\
\rm  Victor Oknyanskij\\
\smallskip
 \it Sternberg Astronomical Institute, Universitetskij pr., 13, Moscow, 119899, Russia\\
\smallskip
\end{center}



 \abstract{We collected from literature  the information about large-separated
 (more then $3^{\prime \prime}$}) pairs of QSOs, which however once
 were suspected as gravitationally lensed system. We discuss some
 new results on time delay determinations including optical-radio correlations
 for QSO 0957+561. We considered some possible observational effects of
 gravitational lensing by a cosmic string. A future international project for observations of
the  gravitational lens system UM425 is briefly discussed.

\section {Introduction}

QSO 0957+561 is not only most famous and wide accepted gravitational
lens system, but also is the prototype of a class of double QSOs with angular
separation larger then, say, 3 arcseconds, the `same' redshift and
`similar' spectra.  If in cases of multiple QSOs with smaller
separations interpretation as a gravitational lens  systems is more
clear, for the class of large angular separation system this task is
more difficult. One of the convincing arguments in favor of
gravitational lens interpretation could be evidence that light variations of
both images are correlated with some time delay. For QSO 0957+561 this
argument was got in result of extensive optical and radio monitoring
during more then 15 year and numerous statistical investigations of the
obtained data. However some problems still saved (Oknyanskij\cite{16}; see
also some details in the next section), correlation of light curves for
A and B images of QSO 0957+561 is revealed and there are no any doubts
in gravitational lens nature of the system, now. Situation for all
other known large separated QSOs is rather unclear and they often called
`dark matter lenses' (DMLs), however some part of them must be
binary QSOs (Schneider\cite{23}).  Both of these possibilities are interesting
to stimulate new observational projects. One of these projects will briefly
discussed below.

\vskip 5cm

\begin{table}[b]
\caption {\bf List of known large separated double QSOs.}
\vskip0.5cm
\begin{center}
\begin{tabular}{|c|c|c|c|c|} \hline
 {\bf Object}&{$\bf \vartheta$}&{\bf z}&{$ \Delta {v (km/s)}$}&{\bf $B_1$:$B_2$}\\
&&&&\\
\hline
0957+561& $6.^{\prime \prime} 1$& 1.41& 0 & 17.5:17.7 \\
\hline
1343+264& $9.^{\prime \prime} 5$& 2.03& 100& 20.2:20.1\\
\hline
2345+007& $7.^{\prime \prime}3$& 2.15 & 15 & 19.5:20.1\\
\hline
1120+019=UM425 & $6.^{\prime \prime} 5$& 1.46& 200& 16.2:20.8 \\
\hline
1429-008& $ 5.^{\prime \prime} 1$ & 2.08 &  260 & 17.7:20.8 \\
\hline
0023+171 & $ 4.^{\prime \prime}8$ & 1.35&  0    & 22.8:23.4 \\
\hline
1145-071& $ 4.^{\prime \prime}2$ & 1.35& 100 & 18:19 \\
\hline
1634+267  & $3.^{\prime \prime}8$ & 1.96 &  150 & 19.2:20.8 \\
\hline
HE 1104-1805 &  $3.^{\prime \prime}0$ & 2.30 &  300 & 16.7:18.6 \\
\hline
Hazard 1146+111  & $2.^{\prime} 6 $& 1.01 & 150 & 18.9:19.5 \\
\hline
\end{tabular}
\end{center}
\end{table}

 \section {Known large-separated double QSOs}
We collected from the literature  informations (Tabl.1)
on the large-separated double QSOs, which were however once suspected as
gravitational lens. Only Q0957+561 is accepted case of gravitational lensing.
It is possible that most of these listed  system is not lens system, but binary
QSOs. There are several cases, when  spectral similarities say rather
on lens origin of the systems. Small spectral difference are possible and
even expected, since expected time delay is about year for $5^{\prime
\prime}$ separation and more than 1000 years if separation is
$2.^{\prime}6$ . Additionally microlensing can change line profiles, as
well as mean redshifts.

{\bf Comments to individual objects in Tabl. 1:}
\smallskip

{\bf 0957+561.} This is the first reported example and the best-known
 lens system. The principal lensing galaxy at z=0.36 is located very
 close (about $1^{\prime \prime}$) to
image B. This main lensing galaxy is situated to the centre of a galaxy
cluster. This cluster and possibly one more cluster at z=0.5 must be taken
into account for the lens model, thereby the existed models still have
large uncertainties.
 The double quasar 0957+561 A,B is up to now only
gravitational lens system for which serious attempts have been made to determine
the time delay $\tau$ between its images, and till now it has been the
most attractive system for this task (see arguments in Beskin \&
 Oknyanskij\cite{3}).
Determination of $\tau$ have a great cosmological interest since may be used to
determine  independently the Hubble constant as well as the age of
Universe (Refsdal\cite{21}).  In despite of intensive attempts to obtain
 correct value of the time delay using long-term optical and radio
 monitoring data sets we have not  some time delay value, which would
 be recognized by all specialists working in this field. Publications
 on the time  delay determination for QSO 0957+561 A,B can be divided
 on four groups due to different obtained results:

1.The time delay value is about 400-425 days (Schild\cite{24},
Schild \& Thomson\cite{25}\cite{19}, Vanderriest et al.\cite{29}
Pelt et al.\cite{18}\cite{19}).

2. The time delay value is about 520-555 days (Beskin \& Oknyanskij
\cite{2}\cite{4};
Press, Rybicki \& Hewitt\cite{18})

3. The time delay value is about 440-455 days (Haarsma et al.\cite{8})

4. The  definite time delay value was not found in view of gaps in data sets
 and possible microlensing effects (Falco et al.\cite{11})

However the value about 425 days is preferred today, we must to note that
some room for other possibilities is still saved. The main reason
for our doubts in value about 425 days are connected with
incorrect methods, which were using to get it. For example, using of
"standard" cross-correlation method based on calculation of direct and
reverse Fourier transforms (Vanderriest et. al.\cite{29}; Schild\cite{24}) is
absolutely incorrect for unequally spaced data (Scargle\cite{22};
Oknyanskij\cite{16}, Beskin \& Oknyanskij\cite{3}). We have got the time
delay value about 410 days with this method if use only real data for B
image, but in place of real measurements for A image take artificial
white noise data and add some line trend.  The question is how probable
 chance coincidence of an incorrectly obtained value with the real one?
If we take into account that the real time delay should be in the
interval 400-600 days and that real accuracy of time delay
determination is about 15 days or worse, then the we can conclude that
probability  for this coincidence is not so small. In our opinion we should
prefer those from several possible values  of the time delay (with about the
same significance) which is farther from the possible artifact value.
Meanwhile only some new independent observational tests could decide
the time delay controversy.  For the time delay value $\tau$ = 420 days
the lens models (see for example, Pelt et al.\cite{19}\cite{20})
the Hubble constant estimated to be smaller than 70 km/(s Mpc). For the
$\tau$=530 days the same estimation gives limit 55 km/(s Mpc).

Radio-optical correlation in the Q0957+561 was first  preliminary
reported by Oknyanskij \& Beskin\cite{15} (here after OB) on the base
of the VLA radio observation during 1979-1990. OB used a  clear  idea  to
take into account the known gravitational lensing time  delay  to
get combined radio and optical light curves and then to use  them
for determination of the possible radio-from-optical time  delay.
It was found this way that  radio  variations  (5  MHz)  followed
optical ones by about 6.4 years with high  level  of  correlation
(~0.87). Using new radio data (Haarsma et al.\cite{8} and take into
account $\tau=425$ days, we have got
for  interval  1979-1994  nearly  the  same    value    of    the
optical-to-radio delay as it had been found before.  Additionally
we suspect that the time delay value linearly increased on  about
120 days per year and intensity of radio-response decreased  with
time.  It is interesting task for future observation project:
try to get the redshift from radio observations of the object. If
it would be obtained then we will have opportunities for
exact estimates of the location and velocity of the variable radio source .
Now, our constrain depend from the unknown orientation of radio region
relative the line of sight. We have made a  conclusion
that the variable radio source is a compact region which is ejected from
 the  central  part of the QSO.  Perhaps,  Q0957+561  is  physically
related to Blazar type objects, but it has different  orientation
relative to the line of sight.

{\bf 1343+264.}
However difference between redshifts of components is much smaller then
measurement errors, there are a strong differences in line profiles and
equivalent widths (Crampton et al.\cite{5}).

{\bf 2345+007.} This object could be called the prototype of
"dark matter lenses". The line profiles is very similar (Steidel and Sargent
1991); however, equivalent widths are different. No possible lens has been
found yet (Tyson et al.\cite{28}). Variability of intensity ratio A/B have been
reported (Weir \& Djorgovski\cite{30}), but regular photometrical monitoring
is not started yet.

{\bf 1120+019=UM425.} Line shapes is similar (Meylan and Djorgovski\cite{14}),
 but variability continuum
in both components are very different. It was suspected that this difference
is connected with microlensing case.  The microlensing hypothesis
was  also pointed out   by Courbin et al.\cite{4}  to explain difference in
light curves of the images.
It is known that A image have {\rm
Broad Absorption Line (BAL)} structure  in O VI $\lambda$ 1033 and
N V $\lambda$ 1240 (Michalitsianos \& Olversen\cite{13}). If it will be found
that B component have also the {\rm BAL} structure, then we will have
very strong argument in favor of gravitational lensing nature of the
 system.

{\bf 1429-008.}  The spectra  of the components are very similar, however
small but significant line profile difference was found (Hewitt et al.\cite{9}).

{\bf 0023+171.} Spectra are similar, but equivalent widths of lines are
different in the components. The components have very complex radio
structure (Hewitt et al.\cite{9}). Perhaps, this system is triplets.

{\bf 1145-071.}  The optical spectra are very similar and all emission
line (excepting C~IV)  have the same equivalent widths
 (Djorgovski et al.\cite{6}). Only the A image is
radio source, but the fainter being at least several hundreds times weaker
in radio wavelengths. This fact can say against gravitational lensing
interpretation, however it can be explained by time delay or existence
of some chance intermediate radio source. If these two opportunities will
be removed any way, then we must involve  some new ideas or
refuse gravitation lensing explanation.

{\bf 1634+267.}  Profiles and continuum shapes are very similar and
support the gravitational lensing interpretation of the system (Steidel
and Sargent\cite{26}). High-ionization lines have velocity difference
up to 1000 km/s, but low-ionization lines have no significant difference.

{\bf HE 1104-1805} Emission line ratios and shapes are very similar, but
continuum spectra are significantly different (Wisotzki et al.\cite{31}).
 Absorption lines are significantly different. The absorption lines in
the two components were intensively investigated with aim of setting
 limits on the sizes of clouds producing the absorption systems (Smette
 et al.\cite{27}). The low-ionization lines are much weaker in the
 spectrum of image B. This difference may be explained by microlensing.

{\bf Hazard 1146+111 B,C.} This object is best example of very large separated
 (more than $1^{\prime}$ ) multiple QSOs with similar redshifts and
  spectra (Arp \& Hazard\cite{1})
We found in literature  5 pairs or triplets of QSOs, which have  very  close
types of spectra, red shifts  and separations about several arc min
(Oknyanskij\cite{17}). It is clear that usual galaxy could not be a lens
for this case, so a cosmic string hypothesis was several times
discussed for this system (see for example, Gott\cite{7}).
        The last few years the  cosmic string hypothesis for explaining
        twin QSOs with several arc  min separations was not used, because
        observations of microwave  background  decreased  the limit  and,
         consequently the possible angular separation of QSOs lensed
         by a cosmic  string.  We  must  to  note  that
        amplitude of microwave background  depend  from  direction  of
        the cosmic string speed and  we  can  admit  existence  of a cosmic
        string with linear density $\mu$ more than value of the CBG
     anisotropy $(\delta T/T)$. The rejection of the Q1146+111 B,C as
	 possible effect of cosmic string lensing have other additional
	 reasons. First of all we must expect several other double
	 images, however one of other D, N, H,  K  QSOs in  the  same  field.
        Note that this  problem can have very simple solution if we take
        into account possible variability of QSOs and different local time for
        the images. Second images perhaps exist, but have  low  for
        observation  intensities  in  present  moment  of  time.  In the case
        of gravitational lensing by
        an Alice cosmic string  we can expect more
        different  brightnesses, redshifts and spectra of double images, because they correspond
        actually to the two different QSOs: from usual  and
        "mirror"  matter (Khlopov \& Sazhin\cite{12};
        Oknyanskij\cite{17}).  That give us some opportunity to use the
	 lensing by a cosmic string  as possible  interpretation  not
	 only for Q1146+111 system for other twin QSOs  with several
	arc min separations.  Meanwhile, we should to state, that in
	the absence of any new positive evidence and in the presence of
	reported differences in the UV and IR spectra for the 1146+111
	B,C the interest to this object as possible gravitational lens
	is weaker now than before.

\section  {How to recognize a cosmic string between large-separated double QSOs?}

1. In case of lensing by cosmic string we must observe only pair of
QSOs images or several pairs along the straight line, but in case
of triplet images some other model of lens must be involved without any
doubts.

2. We must we sure that there are no some other type of lensing objects
(for example, clusters of galaxies).

3. In case of gravitational lensing by a cosmic string the position angle of
polarization must be the same in both images.

4.  If the effect  mentioned above (3) will be found for some system then
it is interesting to search additional images of lensing objects
in  vicinity of the pair of QSOs along the direction determined by the
position angle of polarization.

5. In the Alice string case we can expect variations of the broad line
redshift values with the same period in both images of some lensed QSO
(Oknyanskij\cite{17}).

6. In the Alica string case we can expect
significant difference in brightnesses, spectra and variability of components
(Oknyanskij\cite{17}). So we can expect that some of images along the cosmic string
will have brightness below than the observational limit, and therefore
     only single QSOs could be observed in place of some pairs images.

\section {Observational project}

In collaboration with Drs. Courbin and P.Magain (Liege, Belgium) we are
going to apply for observations with the 6-m telescope of very interesting  object UM425 using
6-m telescope. The following are main goals of the observational
project:

- We wish to obtain deep R and B images of UM 425  in order to combine
the high S/N  final images with HST and  NTT high resolution images in
order to study  the weak deformations  of background galaxies, as well
as to continue the monitoring of this unique lensed quasar.

- We would like to obtain a simultaneous spectrum  of the 2 components
of the lensed  system in  order to  test the  microlensing hypothesis
pointed out   by Courbin et al.\cite{4}, as    done for
HE1104-1805 at La Silla Observatory  (Wisotzki et al.\cite{31}; Smette et
al.\cite{27}).

- We wish to  reach a  limiting  magnitude of  R $\sim$26  and B $\sim$ 27. This   is
feasible in 4 hours of exposure using the 3.5m NTT in La Silla, Chile,
in imaging mode. The same kind of performance can be expected from the
6-m telescope.

\section {Conclusion}
 Ten  to twenty  quasars  are  now known  to   be  multiply  imaged  by
gravitational  lensing.  The study   of these objects  opens important
prospects  in cosmology since  the phenomenon of gravitational lensing
is very rich  in applications: it provides  us with a unique probe  of
the distant Universe and large-scale distribution of matter; a testing
range  for  the theories  of    gravitation;  a  ruler  for  the  size
measurements of intergalactic clouds; the most  sensitive test for the
value of  the  cosmological constant; a   new handle on  the values of
other  cosmological  parameters.    Lensed quasars   should  allow the
determination  of the Hubble constant  and, possibly, the deceleration
parameter,  independently  of the  classical methods  of observational
cosmology.

However, the determination  of these cosmological parameters, which is
based on a  measure of the  time delay  between the different  images,
requires the knowledge of the deflecting  potential.  All the galaxies
(or clusters of galaxies) which contribute to the bending of the light
rays  should be  identified   and  their mass distribution   precisely
estimated.  So  far, in many cases, even  the main  lensing galaxy has
not been detected yet.  Finally we can conclude that some room for
involving of any type exotic `dark matter lens' still saved.


\begin{thebibliography}{1} \itemsep=-5pt

\bibitem{1} Arp, H., Hazard, C. 1980, Ap. J., {\bf 240}, 726.
\bibitem{2}Beskin, G.M. \& Oknyanskij, V.L., 1992, in: Kayzer R., Schramm T.,
Refsdal S. (eds), Lecture Notes in Fysics {\bf 406}, Gravitational Lenses,
Springer, Heidelberg, p.67
\bibitem{3} Beskin, G.M.\& Oknyanskij, V.L., 1995, A\&A, {\bf 304}, 341
\bibitem{4} Courbin, F. et al.  1995,   A\&A, {\bf 303}, 1
\bibitem{5} Crampton, D. et al. 1988, ApJ, {\bf 330}, 184
\bibitem{6} Djorgovski, S.  et al. 1987, ApJ, {bf 321}, L21
\bibitem{7} Gott, J.R. 1985, ApJ, {\bf 288}, 422
\bibitem{8} Haarsma D.B. et al. 1996 in: Kochanek C.S. \& Hewitt J.N. (eds),
Astrophysical Application of Gravitational Lensing, IAU Sym 173, p.43
\bibitem{9} Hewitt, J.N. et al. 1987, ApJ, {\bf 321}, 706
\bibitem{10} Hewitt, P.C. et al. 1989, ApJ, {bf 346}, L61
\bibitem{11} Falko, E.E., Wambsganss, J., \& Schneider P., 1991,  MNRAS, {\bf 251} 698
\bibitem{12} Khlopov, M. Yu., \& Sazhin, M.V. 1989, AZh, {\bf 66}, 191
\bibitem{13} Michalitsianos, A.G. \& Olversen, R.J. 1996, in: Kochanek C.S. \& Hewitt J.N. (eds),
Astrophysical Application of Gravitational Lensing, IAU Sym 173, p. 257
\bibitem{14} Meylan and Djorgovski 1989, ApJ, {\bf 338}, L1
\bibitem{15} Oknyanskij V.L \& Beskin G.M. , 1993 in: Surdej J. et al. (eds), 31st Liege
 International Astrophysical Colloquium: Gravitational lenses
 in the Universe, p.65
\bibitem{16} Oknyanskij V.L., 1996,in: Kochanek C.S. \& Hewitt J.N. (eds),
Astrophysical Application of Gravitational Lensing, IAU Sym 173, p.45
\bibitem{17}  Oknyasnkij, V.L. 1996, in: Khlopov M.Yu. et al. (eds),
"Cosmion-94", Proceeding of international conference, Editions Frantieres, p.321
\bibitem{18}  Press, W.H., Rybicki G.B., \& Hewitt J.N. 1992, ApJ, {\bf 385}, 404
\bibitem{19}  Pelt, J. et al. 1994, A\&A, {\bf 286}, 453
\bibitem{20}  Pelt, J. et al. 1996, A\&A, {\bf 305}, 97
\bibitem{21} Refsdal, S. 1964, MNRAS, {\bf 128}, 307
\bibitem{22}  Scargle J.D. 1989,  Ap.J., {bf 343}, 887
\bibitem{23} Schneider, 1993  in: Surdej J. et al. (eds), 31st Liege
 International Astrophysical Colloquium: Gravitational lenses
 in the Universe, 47
\bibitem{24}  Schild, R.E.1990, AJ, {\bf 100}, 1771
\bibitem{25}  Schild, R.E. \& Thomson, D.J. 1996,in: Kochanek C.S. \& Hewitt J.N. (eds),
Astrophysical Application of Gravitational Lensing, IAU Sym 173, p. 51
\bibitem{26} Steidel, C.C. \& Sargent, W.L.W. 1991, AJ, {\bf 102}, 1610
\bibitem{27} Smette, A.  et al., 1996, in: Kochanek C.S. \& Hewitt J.N. (eds),
Astrophysical Application of Gravitational Lensing, IAU Sym 173, p. 103
\bibitem{28} Tyson, J.A.  et al. 1986, AJ, {\bf 92}, 691
\bibitem{29} Vanderriest, C. et al. 1989, A\&A, {\bf 215}, 1
\bibitem{30} Weir, N., \& Djorgovski, S. 1991, {\bf 101}, 66
\bibitem{31} Wisotzki et al., 1993, A\&A, {\bf 278}, L15
\end{thebibliography}
\end{document}